\documentclass{article}

\usepackage{arxiv}

\usepackage[utf8]{inputenc} % allow utf-8 input
\usepackage[T1]{fontenc}    % use 8-bit T1 fonts
\usepackage{hyperref}       % hyperlinks
\usepackage{url}            % simple URL typesetting
\usepackage{booktabs}       % professional-quality tables
\usepackage{amsfonts}       % blackboard math symbols
\usepackage{nicefrac}       % compact symbols for 1/2, etc.
\usepackage{microtype}      % microtypography
\usepackage{lipsum}

\usepackage{graphicx}        % standard LaTeX graphics tool
\usepackage{multicol}        % used for the two-column index
\usepackage[bottom]{footmisc}% places footnotes at page bottom
\usepackage{url}
\usepackage[utf8]{inputenc}
\usepackage[table]{xcolor}

\title{P2P Loan acceptance and default prediction with Artificial Intelligence}

\author{
  Jeremy D. Turiel \\
  Department of Computer Science\\
  University College London\\
  Gower St, Bloomsbury \\
  London WC1E 6BT, United Kingdom \\
  \texttt{jeremy.turiel.18@ucl.ac.uk} \\
   \And
  Tomaso Aste \thanks{Head of the Financial Computing and Analytics Group \url{http://www.cs.ucl.ac.uk/staff/tomaso_aste/}. Director, UCL Centre for Blockchain Technologies \url{http://blockchain.cs.ucl.ac.uk/tomaso-aste/}.} \\
  Department of Computer Science\\
  University College London \&\\
  Gower St, Bloomsbury \\
  London WC1E 6BT, United Kingdom \\
  \texttt{t.aste@ucl.ac.uk} \\
}

\begin{document}
\maketitle

\begin{abstract}
Logistic Regression and Support Vector Machine algorithms, together with Linear and Non-Linear Deep Neural Networks, are applied to lending data in order to replicate lender acceptance of loans and predict the likelihood of default of issued loans. A two phase model is proposed; the first phase predicts loan rejection, while the second one predicts default risk for approved loans. Logistic Regression was found to be the best performer for the first phase, with test set recall macro score of $77.4 \%$. Deep Neural Networks were applied to the second phase only, were they achieved best performance, with validation set recall score of $72 \%$, for defaults. This shows that AI can improve current credit risk models reducing the default risk of issued loans by as much as $70 \%$. The models were also applied to loans taken for small businesses alone. The first phase of the model performs significantly better when trained on the whole dataset. Instead, the second phase performs significantly better when trained on the small business subset. This suggests a potential discrepancy between how these loans are screened and how they should be analysed in terms of default prediction.
\end{abstract}

% keywords can be removed

\section{Introduction}
\label{sec:1}

Accurate prediction of default risk in lending has been a crucial theme for banks and other lenders for over a century. Modern days availability of large datasets and open source data, together with advances in computational and algorithmic data analytics techniques, have renewed interest in this risk prediction task.
Furthermore, automation of the loan approval process opens new financing opportunities for small businesses and individuals. These previously suffered from more limited access to credit, due to the high cost of human processing. Ultimately, automation of this process carries the potential to reduce human bias and corruption, making access to credit fairer for all. Financial technologies are having a strong impact on this domain, which is rapidly changing \cite{deloittelending}. The application of this model to P2P lending is just one example, with others being micro-financing in developing countries and loan-by-loan evaluation of loan portfolios for investment.

P2P lending has attracted the attention of industry, academics and the general public in recent years. This is also due to the large expansion of major P2P lending platforms like the Lending Club, which has now lent over \$45bn to more than 3mln customers. Another reason for the increasing coverage and popularity of P2P lending is its fast expansion to less developed markets in Eastern Europe, South America and Africa. As the monetary and social relevance of the industry grows, the need for regulation arises. The FCA is among the regulars which have set rules for this industry \cite{ftFCA, FCAP2P}, indicating the importance of the trend in developed countries other than the United States.

Thanks to its easily accessible historical datasets, the Lending Club is the subject of multiple publications investigating the drivers of default in P2P lending \cite{mollenkamp2017determinants, emekter2015evaluating}. The growth of P2P lending in emerging countries has also attracted research interest, for instance \cite{canfield2018determinants} investigates lending in Mexico. This highlights the crucial role of P2P lending in providing access to credit for the population of emerging countries. Interdisciplinary scientific communities such as that of network science have also started to show interest in the socioeconomic dynamics of P2P lending \cite{lin2013judging}. More theoretical works have also inquired about the reason for the need and growth of P2P lending. This was often connected to the concept of credit rationing due to asymmetric information between lending counterparts \cite{stiglitz1981credit}.
A solution to the problem of credit rationing, focused towards allowing fair access to credit and reducing poverty, are micro-finance institutions. Chris Anderson, Editor in Chief of $Wired$ magazine, already identified the concept of ``selling less of more", which is now making its way through to the lending market \cite{anderson2006long}. In order to reduce frictions and allow MFIs to have a self-sustainable business model Serrano-Cinca et al. already suggested that technology will allow to reduce costs and interest rates, leading to an e-commerce like revolution \cite{serrano2014microfinance}. This work aims to contribute to this goal.

To the best of our knowledge, academic publications investigating the drivers of P2P lending \cite{mollenkamp2017determinants, emekter2015evaluating, canfield2018determinants} have applied simple regression models to this task. This work constitutes a significant step forward to applying Big Data and Artificial Intelligence techniques to P2P lending, combining two major disruptive emerging fields.

The rest of the paper is organised as follows: in Section \ref{sec:2} we describe the dataset used for the analysis and the methods applied, in Section \ref{sec:3} we present results and related discussion for the first (Section \ref{subsubsec:3.1.1}) and second phase (Section \ref{subsubsec:3.1.2}) of the model applied to the entire dataset, Section \ref{subsec:3.2} then investigates similar methods applied in the context of ``small business'' loans, Section \ref{sec:4} draws conclusion from our work, followed by acknowledgments in Section \ref{sec:5}.

\section{Dataset and Methods}
\label{sec:2}
% Always give a unique label
% and use \ref{<label>} for cross-references
% and \cite{<label>} for bibliographic references
% use \sectionmark{}
% to alter or adjust the section heading in the running head

\subsection{Dataset}
\label{subsec:2.1}

\begin{figure}[ht]
%\sidecaption
\centering
% For example, with the graphicx style use
\includegraphics[width=120mm]{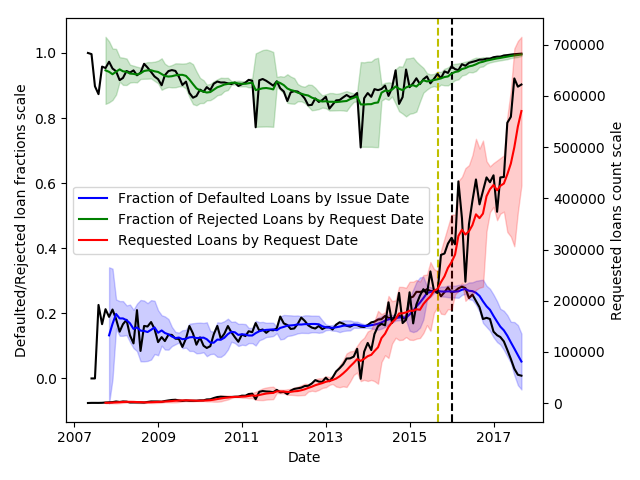}
%
% If no graphics program available, insert a blank space i.e. use
%\picplace{5cm}{2cm} % Give the correct figure height and width in cm
%
\caption{Time series plots of the dataset \cite{data}. Three plots are presented: the number of defaulted loans as a fraction of the total number of accepted loans (blue), the number of rejected loans as a fraction of the total number of loans requested (green) and the total number of requested loans (red). The black lines represent the raw time series, with statistics (fractions and total number) computed per calendar month. The coloured lines represent six-month moving averages and the shaded areas of the corresponding colours represent the standard deviation of the averaged data. The data on the right of the vertical black dotted line was excluded due to the clear decrease in the fraction of defaulted loans, this was argued to be due to the fact that defaults are a stochastic cumulative process and that, with loans of 36-60 months term, most loans issued in that period did not have the time to default yet. A larger fraction of loans is, instead, repaid early. This would have constituted a biased test set.}
\label{fig:1}       % Give a unique label
\end{figure}

The data was collected from loans evaluated by Lending Club in the period between 2007 and 2017 (\url{https://www.lendingclub.com}). The dataset was downloaded from Kaggle (\url{www.kaggle.com}).

%We analyse the two datasets: one contains all the loans which were rejected by analysts of the Lending Club and the other contains all loans which were accepted.
In this paper, we present the analysis of two rich open source datasets \cite{data} reporting loans including credit card-related loans, weddings, house-related loans, loans taken on behalf of small businesses and others. One dataset contains loans that have been rejected by credit analysts, whilst the other, which includes a significantly higher number of features, represents loans which have been accepted and indicates their current status. Our analysis concerns both.
The first dataset comprises over 16 million rejected loans, but has only 9 features. The second dataset comprises over 1.6 million loans and it originally contained 150 features.
We cleaned the datasets and combined them into a unique dataset containing $\approx 15$ million loans, including $\approx 800,000$ accepted loans. Almost $800,000$ accepted loans labelled as ``current" were removed from the dataset, since no default or payment outcome was available.
%Parameter tuning in the grid search was optimised to maximise the unweighted recall average of the two class labels (accepted and rejected).
The dataset of accepted loans indicates the status of each loan. Loans which had a status of fully paid (over $600,000$ loans) or defaulted (over $150,000$ loans) were selected for the analysis and this feature was used as target label for default prediction. The fraction of issued to rejected loans is $\simeq 10\%$, with the fraction of issued loans analysed constituting only $ \approx 50 \%$ of the overall issued loans. Defaulted loans represent $15 - 20 \%$ of the issued loans analysed.

In the present work, features for the first phase were reduced to those shared between the two datasets. For instance, geographical features (U.S. state and postcode) for the loan applicant were excluded, even if they are likely to be informative. Features for the first phase are: 1) debt to Income ratio (of the applicant); 2) employment length (of the applicant); 3) loan amount (of the loan currently requested); 4) purpose for which the loan is taken. In order to simulate realistic results for the validation set, the data was sectioned according to the date associated with the loan. Most recent loans were used as validation set, while earlier loans were used to train the model. This simulates the human process of learning by experience. In order to obtain a common feature for the date of both accepted and rejected loans the issue date (for accepted loans) and the application date (for rejected loans) were assimilated into one date feature. This time-labelling approximation, which is allowed as time sections are only introduced to refine model testing, does not apply to the second phase of the model where all dates correspond to the issue date. All numeric features for both phases were scaled by removing the mean and scaling to unit variance. The scaler is trained on the training set alone and applied to both training and test sets, hence no information about the test set is contained in the scaler which could be leaked to the model.

Features considered for the second phase of the model are: 1) loan amount (of the loan currently requested); 2) term (of the loan currently requested); 3) instalment (of the loan currently requested); 4) employment length (of the applicant); 5) home ownership (of the applicant. Rented, owned or owned with a mortgage on the property); 6) verification status of the income or income source (of the applicant. If this was verified by the Lending Club); 7) purpose for which the loan is taken; 8) Debt to Income ratio (of the applicant); 9) earliest credit line in the record (of the applicant); 10) number of open credit lines (in applicant's credit file); 11) number of derogatory public records (of the applicant); 12) revolving line utilisation rate (the amount of credit the borrower is using relative to all available revolving credit); 13) total number of credit lines (in applicant's credit file); 14) number of mortgage credit lines (in applicant's credit file); 15) number of bankruptcies (in the applicant's public record); 16) logarithm of the applicant's annual income (the logarithm was taken for scaling purposes); 17) FICO score (of the applicant); 18) logarithm of total credit revolving balance (of the applicant).

We first analysed the dataset \cite{data} feature by feature to check for distributions and relevant data imbalances. Features providing information for a restricted part of the dataset (less than $70\%$) were excluded and the missing data was filled by mean imputation. This should not relevantly affect our analysis as the cumulative mean imputation is below $10\%$ of the overall feature data. Furthermore, statistics were calculated for samples of at least 10,000 loans each, so the imputation should not bias the results. A time series representation of statistics on the dataset is shown in Figure \ref{fig:1}.

Differently from other analyses of this dataset (or of earlier versions of it, such as \cite{determinants_p2p}), here for the analysis of defaults we use only features which are known to the lending institution prior to evaluating the loan and issuing it. For instance, some features which were found to be very relevant in other works \cite{determinants_p2p} were excluded for this choice of field. Amongst the most relevant features not being considered here are interest rate and the grade assigned by the analysts of the Lending Club. Indeed, our study aims at finding features which would be relevant in default prediction and loan rejection a priori, for lending institutions. The scoring provided by a credit analyst as well as the interest rate offered by the Lending Club would not, hence, be relevant parameters in our analysis.
%The study aims at finding features which would be relevant in default analysis and prediction a priori for the geographic areas.

\subsection{Methods}
\label{subsec:2.2}

Two machine learning algorithms were applied to both datasets presented in Section \ref{subsec:2.1}: logistic regression with underlying linear kernel and Support Vector Machines (see \cite{lr, svm} for general references on these methodologies). Neural Networks were also applied, but to default prediction only. Neural Networks were applied in the form of a linear classifier (analogous, at least in principle, to logistic regression) and a deep (two hidden layers) neural network \cite{nn}.
%Neural networks were applied to default prediction with the aim to construct more complex features to improve the prediction of a non-trivial outcome such as loan default.

%Machine learning is applied to the dataset discussed in Section \ref{subsec:2.1} in order to form a two-phases model. The first phase aims to reproduce human decisions in accepting or rejecting loan applications, this combines the two datasets (one of accepted and the other of rejected loans). The second phase improves on human decisions (or predicted human decisions). The dataset of accepted loans is analysed in order to predict whether the loan will default or will be fully paid. A higher number of features is available for this dataset alone, although certain features were excluded for the choice of field stated in Section \ref{subsec:2.1}.

%For the second phase of the model, most recent loans were used as the validation set while earlier ones were used to train the model, as for the first phase.

Regularisation techniques were applied to avoid overfitting, L2 regularisation was the most frequently applied, but also L1 regularisation was included in the grid search for LR and SVMs. These were included as mutually exclusive, hence not in the form of an elastic net \cite{l1l2, zou2005regularization}. Initial hyperparameter tuning for the model was performed through extensive grid searches. The ranges for the regularisation parameter $\alpha$ varied, but the widest range was $\alpha = [10^{-5}, 10^{5}]$. Values of $\alpha$ were all powers of 10 with integer exponents. Hyperparameters were determined by the grid search and were manually tuned only in some cases specified in Section \ref{sec:3}. This was done by shifting the parameter range in the grid search or by setting a specific value for the hyperparameter. This was mostly done when there was evidence of overfitting from training and validation set results from the grid search. Class imbalance was mitigated through regularisation as well as by balancing the weights at the time of training of the model itself.
Manual hyperparameter tuning was applied as a consequence of empirical evaluations of the model. Indeed, model evaluations through different measures often suggest that a higher or lower level of regularisation may be optimal, this was then manually incorporated by fixing regularisation parameters or reducing the grid search range. Intuition of the authors about the optimisation task was also applied to prioritise maximisation of a performance measure or balance between different performance measures. Training and validation (or test) sets were used in the analysis. The dataset was split at the beginning in order to prevent information leakage, which might provide the model with information about the test set. The test set then contains future unseen data.%Feature scaling is also trained on the training set alone and applied to the test set, in order to avoid information leakage.

Two metrics were used for result validation, namely recall and AUC. AUC can be interpreted as the probability that a classifier will rank a randomly chosen positive instance higher than a randomly chosen negative one \cite{auc}. This is very relevant to the analysis as credit risk and credit ranking are assessed in relation to other loans as well. The metric extrapolates whether defaulting loans are assigned a higher risk than fully paid loans, on average. Recall is the fraction of loans of a class (such as defaulted or fully paid loans) which are correctly predicted. The standard threshold of $50 \%$ probability, for rounding up or down to one of the binary classes, was applied. This is relevant as it does not test the relative risk assigned to the loans, but the overall risk and the model's confidence in the prediction \cite{recall}.

%Grid searches for Logistic Regression and Support Vector Machines were trained to optimise the regularisation parameter and, occasionally, to choose between L1 and L2 regularisation. \cite{l1l2} 

\section{Results and Discussion}
\label{sec:3}

\subsection{General two phases model for all purpose classes prediction}
\label{subsec:3.1}

\subsubsection{First Phase}
\label{subsubsec:3.1.1}

Logistic regression was applied to the combined datasets. The grid search over hyperparameter values was optimised to maximise the unweighted recall average. The unweighted recall average is referred to as recall macro and is calculated as the average of the recall scores of all classes in the target label. The average is not weighted by the number of counts corresponding to different classes in the target label.
We maximise recall macro in the grid search as maximising AUC led to overfitting the rejected class, which bares most of the weight in the dataset. This is due to AUC weighting accuracy as an average over predictions. This gives more weight to classes which are overrepresented in the training set, a bias that can lead to overfitting.
%to reproduce decisions made by the credit analysts in accepting or rejecting individual loans.
%, where $\alpha$ is the regularisation coefficient for L2 regularisation.

In order to obtain a more complete and representative validation set, the split between training and validation sets was $75\%/25\%$ for the first phase of the model (differently from the $90\%/10\%$ split applied in Section \ref{subsubsec:3.1.2}). This provides $25\%$ of the data for testing, corresponding to approximately two years of data. This indeed constitutes a more complete sample for testing and was observed to yield and more stable reliable results.

The grid search returned an optimal model with $\alpha \simeq 10^{-3}$. The recall macro score for the training set was $\simeq 79.8 \%$. Test set predictions instead returned a recall macro score $\simeq 77.4 \%$ and an AUC score $\simeq 86.5 \%$. Test recall scores were $\simeq 85.7 \%$ for rejected loans and $\simeq 69.1 \%$ for accepted loans.

The same dataset and target label were analysed with Support Vector Machines. Analogously to the grid search for logistic regression, recall macro was maximised. A grid search was applied to tune $\alpha$. Training recall macro was $\simeq 77.5 \%$ while test recall macro was $\simeq 75.2\%$. Individual test recall scores were $\simeq 84.0\%$ for rejected loans and $\simeq 66.5\%$ for accepted ones. Test scores did not vary much, for the feasible range of $\alpha = [10^{-3}, 10^{-5}]$.

%Class imbalance for the target feature in the training set is observed to affect the recall scores for the two classes.
In both regressions, recall scores for accepted loans are lower by $\approx 15 \%$, this is probably due to class imbalance (there is more data for rejected loans). This suggests that more training data would improve this score. From the above results, we observe that a class imbalance of almost $20 \textrm{x}$ affects the model's performance on the underrepresented class. This phenomenon is not particularly worrying in our analysis, though, as the cost of lending to an unworthy borrower is much higher than that of not lending to a worthy one. Still, about $70 \%$ of borrowers classified by the Lending Club as worthy, obtain their loans.

The results for SVMs suggest that polynomial feature engineering would not improve results in this particular analysis. The surprisingly accurate results for logistic regression suggest that credit analysts might be evaluating the data in the features with a linear-like function. This would explain the improvements shown by the second phase, when just a simple model was used for credit screening.

\subsubsection{Second Phase}
\label{subsubsec:3.1.2}

Logistic Regression, Support Vector Machines and Neural Networks were applied to the dataset of accepted loans in order to predict defaults. This is, at least in principle, a much more complex prediction task as more features are involved and the intrinsic nature of the event (default or not) is both probabilistic and stochastic.

Categorical features are also present in this analysis. These were ``hot encoded'' for the first two models, but were excluded from the neural network in this work as the number of columns resulting from the encoding greatly increased training time for the model. We shall investigate neural network models with these categorical features included, in future works.
%but they were just provided as categorical features to the neural network (as the algorithm for this is able to handle categorical variables).

For the second phase, the periods highlighted in Figure \ref{fig:1} were used to split the dataset into training and validation sets (with the last period excluded as per the figure caption). The split for the second phase was of $90\%/10\%$, as more data improves stability of complex models and balanced classes had to be obtained through downsampling for the training set (downsampling was applied as oversampling was observed to cause the model to overfit).

In this phase, the overrepresented class in the dataset (fully paid loans) benefitted from the higher quantity of training data, at least in terms of recall score. In this case the overrepresented class is that of fully paid loans while, as discussed in Section \ref{subsubsec:3.1.1}, we are more concerned with predicting defaulting loans well rather than with misclassifying a fully paid loan.

\subsubsection{Second Phase - Logistic Regression}
\label{subsubsec:3.1.3}

The grid search for logistic regression returned an optimal model with a value of $\alpha \simeq 10^{-2}$. The grid was set to maximise recall macro, as for the models in Section \ref{subsubsec:3.1.1}. Training recall macro score was $\simeq 64.3 \%$ and test AUC and recall macro scores were $69.0\%$ and $63.7\%$, respectively. Individual test recall scores were $63.8\%$ for defaults and $63.6 \%$ for fully paid loans, see Table \ref{tab:10}.
Maximising recall macro indeed yields surprisingly balanced recall scores for the two classes. Maximising AUC did not lead to strong overfitting, differently from what is discussed in Section \ref{subsubsec:3.1.1}. Test scores were lower, both in terms of AUC and recall macro.

\subsubsection{Second Phase - Support Vector Machine}
\label{subsubsec:3.1.4}

Support Vector Machines were also applied to the dataset. The optimal value of $\alpha$ returned by the grid search was $\alpha = 10^{-2}$, the same as for logistic regression in Section \ref{subsubsec:3.1.3}. Scores for the model were, though, worse than those returned by logistic regression. Test AUC was $\simeq 64.3 \%$ and individual test recall scores were $58.7 \%$ for defaulted loans and $65.6 \%$ for fully paid loans, see Table \ref{tab:10}. It can be inferred that the analysis of this dataset does not benefit from SVM kernel's non-linearities in its test set performance. Furthermore, recall scores are improved for the overrepresented class in the dataset. This is the opposite of what is aimed for in this analysis, where we prioritise high recall on the default class which has a higher impact on the borrower's balance sheet. Such a strong score imbalance is also not ideal in terms of quality of the predictor. It should be noted that the label class imbalance (defaulted and fully paid loans) is much weaker than that described in Section \ref{subsubsec:3.1.1}, with defaulted loans representing $15-20\%$ of the dataset.

\begin{table}[ht]

\caption{Table with main results from LR and SVM tested for the second phase of the model.}
\label{tab:10}       % Give a unique label
%
% Follow this input for your own table layout
%
\begin{minipage}{\textwidth}
\centering
\begin{tabular}{p{1.5cm}p{1.5cm}p{2cm}p{2cm}p{3cm}}
\hline
\rowcolor{lightgray} \multicolumn{5}{|c|}{Loan Default Prediction Results} \\
\hline\noalign{\smallskip}
Model & $\alpha$ & Recall Train & AUC Test & Recall Test (Macro/Default/Paid) \\
\noalign{\smallskip}

LR & $10^{-2}$ & $ 64.3 \% $ & \cellcolor[HTML]{8AFF33} $69 \%$ & $ 63.7 \% / 63.8 \% / 63.6 \%$ \\
%LR & AUC & $10^{-2}$ & $ 64.3 \% $ & $69 \%$ & $ 63.7 \% / 63.8 \% / 63.6 \%$ \\
SVM & $10^{-2}$ & $ - $ & $64.3 \%$ & $ 62.2 \% / 58.7 \% / 65.6 \%$ \\

\noalign{\smallskip}\hline\noalign{\smallskip}
\end{tabular}

\end{minipage}

%$^a$ Table foot note (with superscript)
\end{table}

\subsubsection{Second Phase - Neural Network}
\label{subsubsec:3.1.5}

Linear Neural Network classifiers as well as Deep (two hidden layers) Neural Networks were also trained on the dataset for the second phase of the model. Linear Neural Network classifiers were trained on numerical features alone as well as on both numerical and categorical features. L2 regularisation was then applied.
Numerical features-only test scores returned an AUC of $67.8 \%$ and a recall of $60.0 \%$ (for defaulted loans). The model yielded improved results when trained on categorical features too. Test scores returned an AUC of $68.7\%$ and recall of $62.7 \%$ (for defaulted loans). These scores are slightly worse than those for logistic regression, but they do not implement regularisation yet. Once L2 regularisation ($\alpha = 10$) was manually set and applied, test AUC improved to $69 \%$ and recall improved to $65 \%$ (for defaulted loans).

A Deep Neural Network (with an arbitrary two hidden layers node structure - $DNN^a$ in Table \ref{tab:9}) was initially applied to numerical data alone. In comparison with the Linear Classifier, test AUC and recall (for defaulted loans) scores improved to $68 \%$ and $67 \%$, respectively. This indeed shows how more advanced feature combinations improve the predictive capabilities of the model. The improvement was expected, as the complexity of the phenomenon described by the target label surely implies more elaborated features and feature combinations than those originally provided to the model.

The DNN was then refined with a grid search on node numbers $n_1, n_2$ for the two hidden layers. The grid search was run over all combinations of values from the sets $n_1 \in \{ 5, 10, 15, 20, 30 \}$, $n_2 \in \{ 1, 3, 5, 10 \}$ and by applying a high level of dropout regularisation ($20 \%$). The level of dropout regularisation was empirically chosen from a $[0 \%, 30 \%]$ range, this is a reasonable range for this type of models often found in the literature. The strong regularisation aimed to reduce the DNN's intrinsic tendency to overfit, leading to a more robust and general model infrastructure. Results on the test set were indeed verified to be largely in line throughout the grid search, suggesting a model which is robust in the context of hyperparameter tuning.

Results for two network structures selected from the grid search (together with $DNN^a$ - arbitrary two hidden layers node structure) are described in Table \ref{tab:9}. These network structures are selected, as their results display the desirable properties of stable AUC and high recall on defaults.

\begin{table}[ht]

\caption{Table with main results from DNN architectures tested for the second phase of the model.}
\label{tab:9}       % Give a unique label
%
% Follow this input for your own table layout
%
\begin{minipage}{\textwidth}
\centering
\begin{tabular}{p{1.5cm}p{1.5cm}p{2cm}p{2cm}p{2.5cm}}
\hline
\rowcolor{lightgray} \multicolumn{5}{|c|}{Loan Default Prediction Results} \\
\hline\noalign{\smallskip}
Model & Dropout & Recall Train& AUC Test & Recall Default Test \\
\noalign{\smallskip}

DNN \footnote{DNN with arbitrary node numbers $[n_1 = 20, n_2 = 5]$} & $20 \%$ & - & $68 \%$ & $ 67 \% $ \\
DNN \footnote{DNN with node numbers fine-tuned to $[n_1 = 30, n_2 = 1]$} & $20 \%$ & $ 71 \% $ & $66 \%$  & \cellcolor[HTML]{8AFF33} $ 75 \% $ \\
DNN \footnote{DNN with node numbers fine-tuned to $[n_1 = 5, n_2 = 3]$}  & $20 \%$ & $ 68 \% $ & \cellcolor[HTML]{8AFF33} $69 \%$ & $ 72 \% $ \\

\noalign{\smallskip}\hline\noalign{\smallskip}
\end{tabular}

\end{minipage}

%$^a$ Table foot note (with superscript)
\end{table}

\begin{figure}[ht]
%\sidecaption
% For example, with the graphicx style use
\centering
\includegraphics[width=120mm]{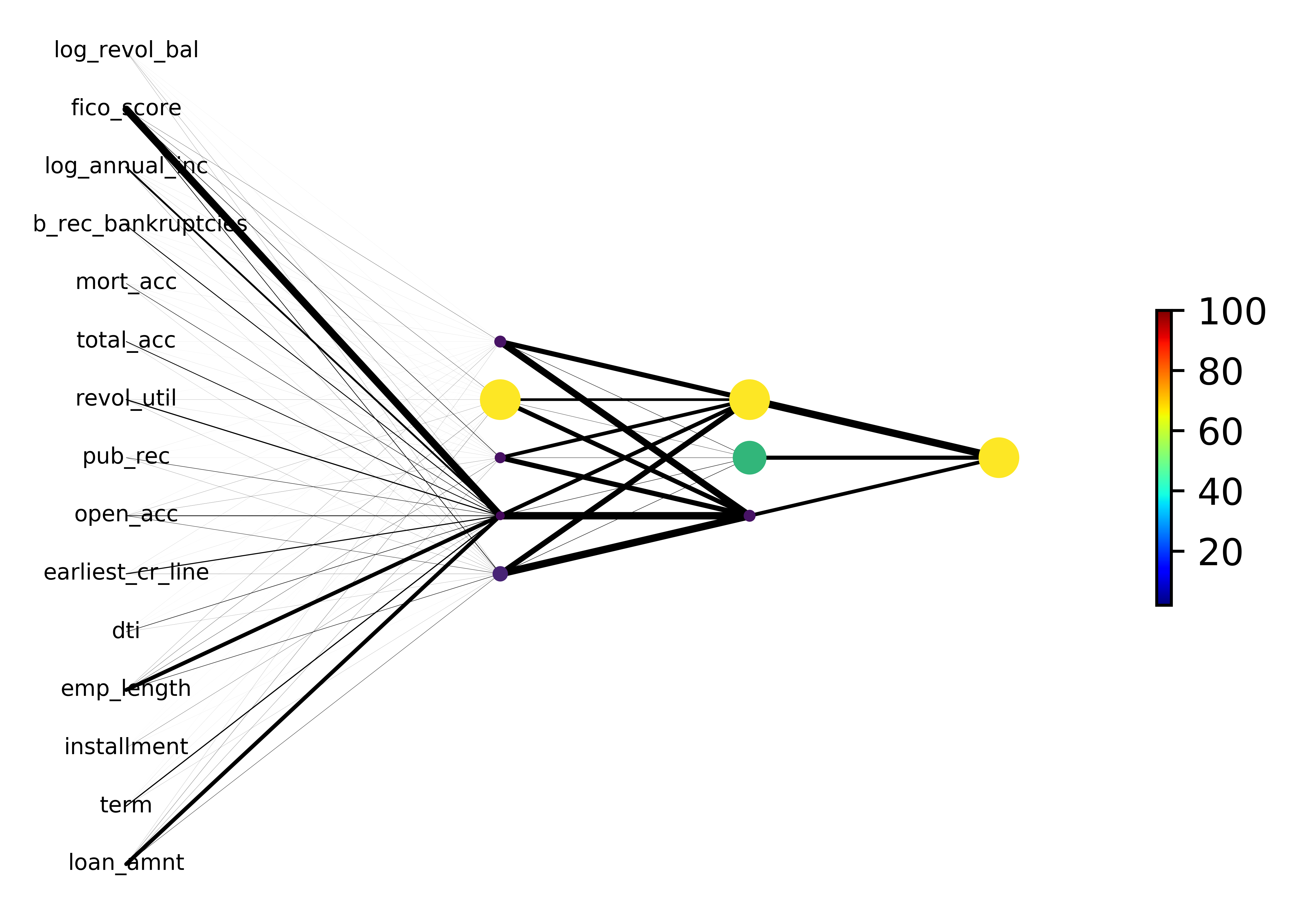}
%
% If no graphics program available, insert a blank space i.e. use
%\picplace{5cm}{2cm} % Give the correct figure height and width in cm
%
\caption{Neural network representation with node size and colour representing total outgoing weight and edge width proportional to the weight. The DNN represented is with node numbers fine-tuned to [5, 3] and $tanh$ non-linearities.}
\label{fig:5}       % Give a unique label
\end{figure}

Figure \ref{fig:5} is a representation of one of the weight instances of the fully trained $DNN^c$ network in Table \ref{tab:9}. The network representation in Figure \ref{fig:5} encodes the weight of each link in the fully connected layer as line thickness. Node size and colour are indicative of the normalised sum of outgoing weights from the node. This representation clearly constitutes an approximation, as the nodes contain non-linearities, but it still provides a useful visual interpretation and stability check tool.

%In order to ensure model interpretability, we have applied a parsimonious approach to modelling when working with Neural Networks. This involves drastically reducing the feature space and deploying a network structure with few hidden layers and nodes per layer. This allows to create a meaningful and insightful representation of the network structure, such as the one presented in Figure \ref{fig:5}. This is a representation of one of the weight instances of the fully trained network with node number fine-tuned to [5, 3], this corresponds to the network described in the last row of Table \ref{tab:9}. The network representation in Figure \ref{fig:5} represents the weight of each link in the fully connected layer by line thickness. Node size and colour are also indicative of the normalised sum of outgoing weights from the node. This representation clearly constitutes an approximation, as the nodes contain non-linearities, but it still provides a useful visual interpretation and stability check tool.

\subsection{Two phases analysis for ``small business" category}
\label{subsec:3.2}

%`` and ''
The ``purpose'' feature described in Section \ref{subsec:2.2} provides information about the purpose for which the loan was requested. The small business class of this feature is of particular interest here. This loan category was observed to have the highest fraction of defaulted loans amongst all categories and the least likelihood to survive throughout the lending term period \cite{determinants_p2p}. Furthermore, this purpose is arguably different from the others and is more business-focused, rather than just a personal loan.

We therefore decided to look at this category in isolation, although it was included in the entire dataset used for the analyses described in the previous sections.

\subsubsection{First Phase - Small Business Training Data Only}
\label{subsubsec:3.2.1}

\begin{table}[ht]
 \caption{Small business loan acceptance results and parameters for SVM and LR grids trained and tested on the data’s “small business” subset.}
  \centering
  \begin{tabular}{lllllll}
    \toprule
    %\multicolumn{2}{c}{Part}                   \\
    %\cmidrule(r){1-7}
    Model & Grid metric & $\alpha$ & Training Score & AUC Test & Recall Rejected & Recall Accepted \\
    \midrule
    LR & AUC  & 0.1 & $88.9 \%$ & $65.7 \%$ & $48.5 \%$ & $62.9 \%$\\
    LR & recall macro & 0.1 & $78.5 \%$ & $65.5 \%$ & $48.6 \%$ & $57.0 \%$\\
    SVM & recall macro & 0.01 & - & $89.3 \%$ & $47.8 \%$ & $62.9 \%$ \\
    SVM & AUC & 10 & - & $83.6 \%$ & $46.4 \%$ & $76.1 \%$ \\
    \bottomrule
  \end{tabular}
  \label{tab:4}
\end{table}

Logistic Regression and Support Vector Machines were trained and tested on ``small business'' loans alone. Two grid searches were trained for Logistic Regression, one maximises AUC whilst the other maximises recall macro. The former returns an optimal model with $\alpha = 0.1$, training AUC score $\simeq88.9 \%$ and test AUC score $\simeq65.7\%$. Individual recall scores are $\simeq48.0 \%$ for rejected loans and $62.9 \%$ for accepted loans. The discrepancy between the training and test AUC scores indicates overfitting to the data or the inability of the model to generalise to new data for this subset. The latter grid search returns results which somewhat resemble the former one. Training recall macro is $\simeq 78.5 \%$ whilst test recall macro is $\simeq 52.8\%$. AUC test score is $65.5 \%$ and individual test recall scores are $48.6 \%$ for rejected loans and $57.0 \%$ for accepted loans. This grid's results again show overfitting and the inability of the model to generalise. Both grids show a counterintuitively higher recall score for the underrepresented class in the dataset (accepted loans) whilst rejected loans are predicted with recall lower than $50 \%$, worse than random guessing. This might simply suggest that the model is unable to predict for this dataset or that the dataset does not present a clear enough pattern or signal.

Support Vector Machines perform poorly on the dataset in a similar fashion to Logistic Regression. Two grid optimisations are performed here too, in order to maximise AUC and recall macro, respectively. The former returns a test AUC score of $89.3 \%$ and individual recall scores of $47.8 \%$ for rejected loans and $62.9 \%$ for accepted loans. The latter grid returns a test AUC score of $83.6 \%$ with individual recall scores of $46.4 \%$ for rejected loans and $76.1\%$ for accepted loans (this grid actually selected an optimal model with weak L1 regularisation). A final model was fitted, where the regularisation type (L2 regularisation) was fixed by the user and the range of the regularisation parameter was shifted to lower values in order to reduce underfitting of the model. The grid was set to maximise recall macro. This yielded an almost unaltered AUC test value of $\simeq 82.2 \%$ and individual recall values of $47.3 \%$ for rejected loans and $70.9 \%$ for accepted loans. These are slightly more balanced recall values. However, the model is still clearly unable to classify the data well, this suggests that other means of evaluation or features could have been used by the credit analysts to evaluate the loans. The hypothesis is reinforced by the discrepancy of these results with those described in Section \ref{subsubsec:3.1.1} for the whole dataset. It should be noticed, though, that the data for small business loans includes a much lower number of samples than that described in Section \ref{subsubsec:3.1.1}, with less than $3 \cdot 10^5$ loans and just $\approx 10^4$ accepted loans.

\subsubsection{First Phase - All Training Data}
\label{subsubsec:3.2.3}

Given the poor performance of the models trained on the small business dataset and in order to leverage the large amount of data in the main dataset and its potential to generalise to new data and to subsets of its data, Logistic Regression and Support Vector Machines were trained on the whole dataset and tested on a subset of the small business dataset (the most recent loans, as by the methodology described in Section \ref{subsec:2.2}). This analysis yields significantly better results, when compared to those discussed in Section \ref{subsubsec:3.2.1}. Results are presented in Table \ref{tab:1}.

\begin{table}[ht]
 \caption{Small business loan acceptance results and parameters for SVM and LR grids trained on the entire dataset and tested on its ``small business'' subset.}
  \centering
  \begin{tabular}{lllllll}
    \toprule
    %\multicolumn{2}{c}{Part}                   \\
    %\cmidrule(r){1-7}
    Model & Grid metric & $\alpha$ & Training Score & AUC Test & Recall Rejected & Recall Accepted \\
    \midrule
    LR & AUC  & 1 & $89.0 \%$ & $71.9 \%$ & $53.5 \%$ & $60.2 \%$\\
    LR & recall macro & 0.1 & $77.9 \%$ & $71.7 \%$ & $54.0 \%$ & $59.9 \%$\\
    LR & fixed  & 0.001 & $80.0 \%$ & $71.1 \%$ & $55.2 \%$ & $65.2 \%$\\
    LR & fixed & 0.0001 & $80.1 \%$ & $71.0 \%$ & $55.9 \%$ & $62.9 \%$\\
    SVM & recall macro & 0.01 & - & $77.5 \%$ & $52.6 \%$ & $68.4 \%$ \\
    SVM & AUC & 10 & - & $89.0 \%$ & $97.3 \%$ & $43.3 \%$ \\
    \bottomrule
  \end{tabular}
  \label{tab:1}
\end{table}

The results presented in Table \ref{tab:1} for Logistic Regression still present consistently higher recall for accepted loans. There is an apparent credit analyst decision bias towards rejecting small business loans. This could, though, be explained as small business loans have a higher likelihood of default, hence they are considered more risky and the model, trained on all the data, does not have this information. Information on loan defaults is present as a label only in default analysis, as no data is present for rejected loans. Future works might input the percentage of defaulted loans corresponding to the loan purpose as a new feature and verify whether this improves the model.

Results for Support Vector Machines are in line with those for logistic regression. The grid trained to maximise AUC is clearly overfitting the rejected class to maximise AUC and should be discarded. Results for the grid maximising recall macro follow the same trend of those from Logistic Regression. Recall scores are slightly more unbalanced. This confirms the better performance of Logistic Regression for the prediction task, as discussed in Section \ref{subsubsec:3.1.1}.

%Results for Support Vector Machines in Table \ref{tab:1} are now different from those for Logistic Regression in the same table. The AUC scores are significantly lower ($\approx 15 \%$) than those discussed for SVMs trained on small business data only, but recall scores are now in line with what is expected, the overrepresented class (rejected loans) is now well predicted. This is also in line with the priorities for the model discussed in the previous sections. Recall scores for accepted loans are, though, below $50\%$. This might suggest that overfitting of the model is the cause, but the difference in recall scores between SVMs and LR shows that the two models indeed develop different approached to the prediction task. Overall, Logistic Regression seems to predict both classes better and indeed to have a higher AUC score too.

\subsubsection{Second Phase}
\label{subsubsec:3.2.4}
%`` and ''
Logistic regression and Support Vector Machines were trained on accepted loan data in order to predict defaults of loans with ``small business'' purpose. Analogously to the analysis discussed in Section \ref{subsubsec:3.2.1}, the models were trained and tested on small business data alone.
Results for models trained on small business data alone are presented in Table \ref{tab:2}. Results for Logistic Regression are slightly worse and more unbalanced in individual recall scores than those presented in Section \ref{subsubsec:3.1.2}, this can be explained by the smaller training dataset (although more specific, hence with less noise). Surprisingly, again, the underrepresented class of defaulted loans is better predicted. This could be due to the significant decay of loan survival with time for small business loans, this data is obviously not provided to the model, hence the model might classify as defaulting, loans which might have defaulted with a longer term. Alternatively, most defaulting loans could be at high risk, while not all risky loans necessarily default, hence giving the score imbalance. Maximising AUC in the grid search yields best and most balanced results for Logistic Regression in this case. Analogously to  the analysis in Section \ref{subsubsec:3.2.1} class imbalance is strong here, defaulted loans are $\approx 3 \%$ of the dataset. The better predictive capability on the underrepresented class might be due to loan survival with time and should be investigated in further works. Three threshold bands might improve results, where stronger predictions only are evaluated.

Support Vector Machines provide more balanced results, although worse overall, for this task. In both SVMs and LR we observe how stronger regularisation, corresponding to higher values of $\alpha$, improves recall results on the test set for the overrepresented class. AUC test scores improve as well, suggesting an improvement in the model's ability to generalise.

\begin{table}[ht]
 \caption{Small business loan default results and parameters for SVM and LR grids trained and tested on the data's ``small business'' subset.}
  \centering
  \begin{tabular}{lllllll}
    \toprule
    %\multicolumn{2}{c}{Part}                   \\
    %\cmidrule(r){1-7}
    Model & Grid metric & $\alpha$ & Training Score & AUC Test & Recall Defaulted & Recall Paid \\
    \midrule
    LR & AUC  & 0.1 & $64.8 \%$ & $66.4 \%$ & $65.2 \%$ & $57.4 \%$\\
    LR & recall macro & 0.01 & $60.4 \%$ & $65.3 \%$ & $64.6 \%$ & $53.3 \%$\\
    SVM & recall macro & 0.01 & - & $59.9 \%$ & $59.8 \%$ & $58.8 \%$ \\
    SVM & AUC & 0.1 & - & $64.2 \%$ & $50.8 \%$ & $65.8 \%$ \\
    \bottomrule
  \end{tabular}
  \label{tab:2}
\end{table}

Analogously to the analysis presented in Section \ref{subsubsec:3.2.3}, Logistic Regression and Support Vector Machines were also trained on all the data and tested on small business data only, in order to leverage the larger datasets, which might share signals with its ``small business'' subset. Results in this case, differ from those in Section \ref{subsubsec:3.2.3}, where an improvement was observed. Results are presented in Table \ref{tab:3}. The model poorly predicts fully paid loans, with a recall score even below $50 \%$. This might suggest that the way these loans are screened is similar to that of other categories, but their intrinsic default risk is very different indeed. This is also observed in the discrepancy in loan survival between these loans and all other loan categories. \cite{determinants_p2p} The optimal parameters returned by the grid suggest weaker regularisation than that for results in Table \ref{tab:2}. For predicting a subset of its data, stronger regularisation might improve results, this could be verified in future works. It should be considered, though, that regularisation might reduce the importance of a small subset of the data, such as that of small business loans. The fraction of the small business subset with respect to the complete dataset is roughly the same for loan acceptance ($\simeq 1.3 \%$) and loan default prediction ($\simeq 1.25\%$). This indeed suggests a difference in the underlying risk of the loan and its factors.

As the conclusions about model generalisation described in Section \ref{sec:4} can be drawn already by comparing LR and SVM models, DNNs are not considered for to the small business dataset analysis in Section \ref{subsec:3.2}. DNNs are considered only for the purpose of improving model performance through more complex models and feature combinations, which is the theme of Section \ref{subsec:3.1}.

\begin{table}[ht]
 \caption{Small business loan default results and parameters for SVM and LR grids trained on the entire dataset and tested on its ``small business'' subset.}
  \centering
  \begin{tabular}{lllllll}
    \toprule
    %\multicolumn{2}{c}{Part}                   \\
    %\cmidrule(r){1-7}
    Model & Grid metric & $\alpha$ & Training Score & AUC Test & Recall Defaulted & Recall Paid \\
    \midrule
    LR & AUC  & 0.001 (L1) & $69.8 \%$ & $68.9 \%$ & $81.0 \%$ & $43.3 \%$\\
    LR & AUC & 0.001 & $69.7 \%$ & $69.2 \%$ & $86.4 \%$ & $35.0 \%$\\
    LR & recall macro  & 0.001 & $64.2 \%$ & $69.2 \%$ & $86.4 \%$ & $35.0 \%$\\
    SVM & recall macro & 0.001 & - & $64.1 \%$ & $77.7 \%$ & $48.3 \%$ \\
    SVM & AUC & 0.001 & - & $69.7 \%$ & $77.7 \%$ & $48.3 \%$ \\
    \bottomrule
  \end{tabular}
  \label{tab:3}
\end{table}

\section{Conclusions}
\label{sec:4}

In this paper we demonstrate that P2P loan acceptance and default can be predicted in an automated way with results above $\simeq 85 \%$ (rejection recall) for loan acceptance and above $\simeq 75 \%$ (default recall) for loan default. Given that the present loan screening has a resulting fraction of default around $20\%$ (see Figure \ref{fig:1}) we can infer that potentially the methodology presented in this paper could reduce the defaulting loans to $10\%$ with positive consequences for the efficiency
of this market. The best performing tools were Logistic Regression for loan acceptance and Deep Neural Networks for loan default. The high recall obtained with linear models on replicating traditional loan screening suggests that there is significant room for improvement in this phase as well.

The loan grade and interest rate features were found to be the most relevant for predicting loan default in \cite{determinants_p2p}. The current model tries to predict default without biased data from credit analysts' grade and assigned interest rate, hence these features are excluded. The Deep Neural Network and Logistic Regression models provide substantial improvements on traditional credit screening. A recall score significantly and robustly above $70 \%$, with AUC scores $\simeq 70 \%$ for the Deep Neural Network, improves even on the logistic regression in \cite{determinants_p2p}. The features provided to the model in our  study generalise to any lending activity and institution, beyond P2P lending. The present work could therefore be augmented in order to predict loan default risk without the need for human credit screening.

%The Deep Neural Network and the Logistic Regression models provide substantial improvements on traditional credit screening. This is achieved in our study by including only features which can be generalised to any lending activity and institution, beyond P2P lending. A recall score significantly and robustly above $70 \%$, with AUC scores $\simeq 70 \%$ for the Deep Neural Network, improves even on the logistic regression where features such as assigned loan grade and interest rate are provided to the model \cite{determinants_p2p}. These features were found in \cite{determinants_p2p} to be the most relevant for predicting loan default. The current model tries to predict default without biased data from credit analysts' grade and assigned interest rate. This could be improved further in order to predict loan default risk without the need for human credit screening.

%Results for the method and dataset, described in Section \ref{sec:2}, were presented in Section \ref{sec:3}.
The two phases model for all loan purposes described in Section \ref{subsec:3.1} showed better performance overall, with well-balanced individual test recall scores for the second phase of $75\%$ for defaulted loans. This shows the ability to predict well above $50\%$ of defaults on loans screened and accepted by credit analysts, while not penalising excessively the acceptance of well performing loans. Training on the whole dataset for the first phase resulted in higher scores when applied to small business loans than when trained on small business loans alone. The opposite was true for the second phase, where default prediction was significantly better overall, when trained on small business loans alone. This suggests a discrepancy between how credit analysts treat these loans and how they might be treated more efficiently, in terms of their default risk and characteristics. Neural Networks were shown to significantly outperform the other models, suggesting that they might be used for default prediction, further to credit analyst screening. Neural Networks could also be combined with Logistic Regression in a conservative model, in order to mitigate their complex and not well-predictable nature. This and further data preprocessing and augmentation should be the subject of further work. We shall further extend our work to areas such as micro-financing in developing countries and loan-by-loan evaluation of loan portfolios for investment as well as to traditional lending. The integration of the present model with predictive modelling based on information filtering network techniques \cite{Tumminello26072005}, \cite{marcaccioli2019polya}, \cite{massara2016network}, \cite{mantegna1999hierarchical} will also be the subject of future research.

\section{Acknowledgments}
\label{sec:5}
The authors acknowledge the EC Horizon 2020 FIN-Tech project for partial support and useful opportunities for discussion. JT acknowledges support from EPSRC (EP/L015129/1). JT acknowledges Dr. Guido Germano for useful feedback and discussions. TA acknowledges support from ESRC (ES/K002309/1),  EPSRC (EP/P031730/1) and EC (H2020-ICT-2018-2 825215).

%Data is freely available from Kaggle: \url{https://www.kaggle.com/wordsforthewise/lending-club}. All interested parties will be able to obtain the data in the same manner the authors did.

\bibliographystyle{unsrt}  
\bibliography{P2PAI}  %%% Remove comment to use the external .bib file (using bibtex).
\end{document}